\documentclass[]{revtex4}

\usepackage{graphicx}
\usepackage{amssymb}

\begin{document}

\title{An Approach to Exascale Visualization: \\
Interactive Viewing of In-Situ Visualization}

\author{Akira Kageyama}
\email{kage@cs.kobe-u.ac.jp}
\author{Tomoki Yamada}
\address{Graduate School of System Informatics, Kobe University, Japan}

\begin{abstract}
In the coming era of exascale supercomputing,
in-situ visualization will be 
a crucial approach
for reducing the output data size.
A problem of in-situ visualization is that
it loses interactivity if a steering method is not adopted.
In this paper, we propose a new method for the 
interactive analysis of in-situ visualization images produced by a batch simulation job.
A key idea is to apply numerous (thousands to millions) in-situ 
visualizations simultaneously.
The 
viewer then analyzes
the image database interactively during postprocessing.
If each movie can be compressed to 100~MB,
one million movies will only require 100~TB, 
which is smaller than the size of the raw numerical data in exascale supercomputing.
We performed a feasibility study using the proposed method.
Multiple movie files were produced by a simulation and
they were analyzed using a specially designed movie player.
The user could  change the viewing angle, the
visualization method, and the parameters interactively
by retrieving an appropriate sequence of images from the movie dataset.
\end{abstract}

\maketitle
 
%------------------------------------
\section{Introduction}
%------------------------------------

In supercomputer simulations, postprocessing has become a bottleneck
during the overall research cycle.
This is because  the disk I/O and the network bandwidths
cannot keep pace with the exponential growth in the computer processor speed.
To resolve this imbalance, the user has to apply a
data reduction method to the
output data in supercomputer simulations.
One approach is to compress the raw numerical data directly.
Discrete Fourier transforms and wavelet transforms are often used for this purpose.
Another approach to output data reduction is to apply the visualization during runtime.
This approach, which is known as in-situ visualization, is effective for reducing the data size because
images are two-dimensional.

In-situ visualization poses new challenges for visualization researchers.
New difficulties arise because today's supercomputers are massively parallel machines.
The target data for visualization are divided into pieces, which
are distributed in network-connected memories.
Thus, new visualization methods and algorithms need to be
implemented on supercomputers for in-situ visualization.
They require good parallel scaling to avoid
degradation of the simulation's scale.
These challenges have focused the attention of the High Performance 
Computing (HPC) community on in-situ visualization.
Recent studies of in-situ visualization for HPC
are reviewed in Section~\ref{sec:related work}.

A key point of in-situ visualization is the correct settings for 
two types of parameters:
visualization parameters and camera parameters.
Visualization parameters include 
applied visualization methods and 
their states and values.
The camera parameters include 
the viewpoint position, direction, field of view, 
and viewing frustum culling.

%---------------------------
 \begin{figure}[htbp]
  \begin{center}
   \includegraphics[width=0.8\linewidth]{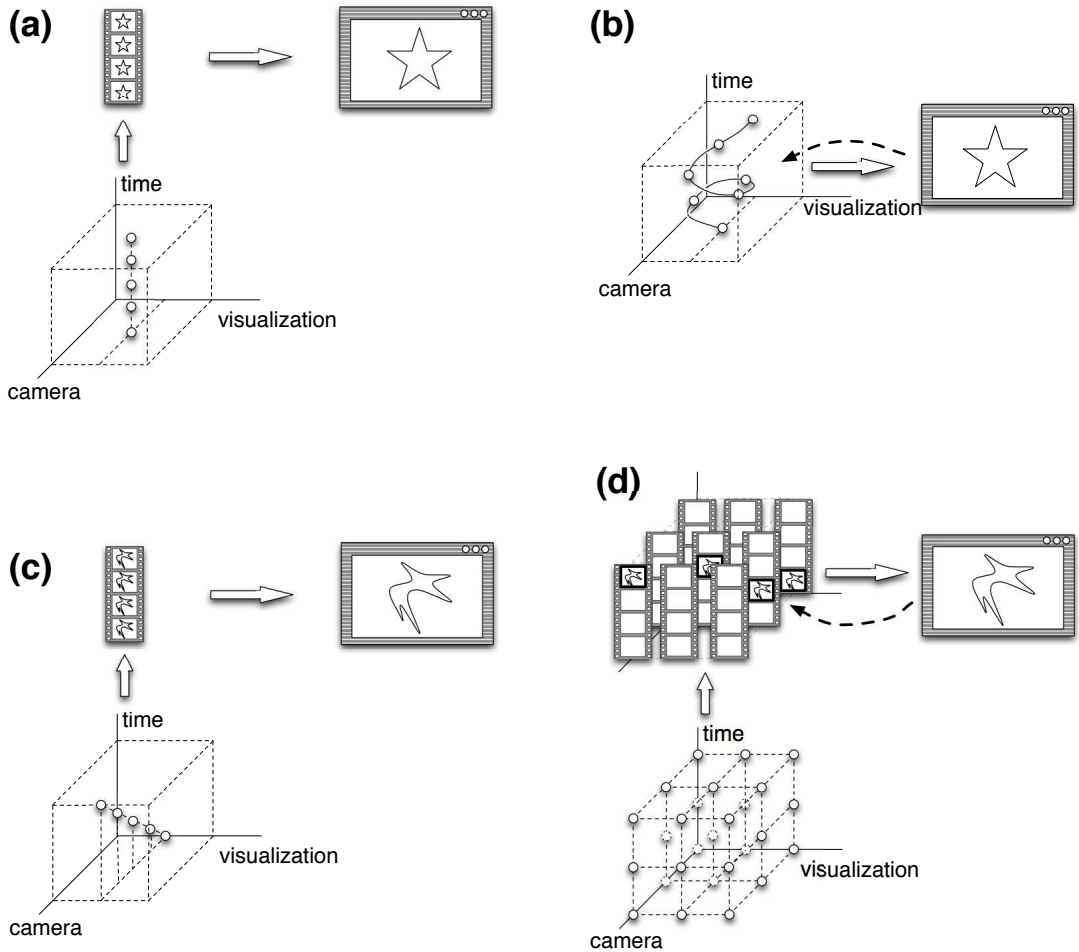}
  \end{center}
  \caption{Concepts used by various visualization approaches.
  (a) Postprocessing.
  The camera-axis shows the viewing freedom, such as
  the viewpoint positions and angles.
  The visualization-axis shows the 
  visualization methods and their internal parameters.
  The third axis shows the simulation time.
  A ball in this camera-visualization-time space represents a visualization shot.
  (b) Steering simulation and visualization.
  (c) Bullet-time method.
  (d) Steering visualization of the in-situ visualization images proposed in this paper.}
  \label{fig:concept_01_04_combined}
 \end{figure}
%---------------------------
A simple in-situ visualization with one fixed camera but 
without steering is described conceptually in
Fig.~\ref{fig:concept_01_04_combined}a,
where a three-dimensional (3-D) space is spanned by 
three bases: the camera parameters,
the visualization parameters, and the simulation time.
The white ball in this figure represents
a visualization shot taken in the 3-D space.
After the simulation task is finished, the 
visualized images are sent to
a local disk drive on the data analyzer's PC.
The white arrow in the figure indicates data transfer.
After this in-situ visualization,
the sequential images are shown as a movie on the PC's display.

In most simulations, the key phenomena or events usually appear 
in localized, discrete spots in the whole simulation region.
If this type of ``hotspot'' is next to the range of the camera view,
the user has to reset the camera positions in the so-called
scenario file and resubmit a simulation job,
unless the steering simulation approach is taken.

In the steering simulation, 
the analyzer can control
the simulation and the visualization parameters dynamically during runtime.
A steering simulation with an in-situ visualization is illustrated in
 Fig~\ref{fig:concept_01_04_combined}b.
The visualized images are transferred to the analyzer during the runtime
of the simulation; 
see the white arrow in Fig.~\ref{fig:concept_01_04_combined}b.
By observing the images on the PC monitor,
the analyzer can change
the visualization parameters dynamically to focus on a hotspot.
The dashed line with an arrow in Fig.~\ref{fig:concept_01_04_combined}b
represents the feedback control from the analyzer to the simulation and visualization.
Although this kind of the steering simulation 
is an effective approach for HPC,
it is not always possible
because supercomputer simulations are 
usually performed in batch jobs.

Without steering, the in-situ visualization loses interactivity completely,
which is critically important for obtaining insights from the simulation.

This paper proposes a method that facilitates
the interactive viewing of the visualization results generated by the in-situ visualization
of a batch-job supercomputer simulation.

%------------------------------------
\section{Interactive View of In-Situ Visualization Data}
%------------------------------------

The key concept used by the method
proposed in this paper
is to produce a cluster of in-situ 
visualizations with different visualization and camera parameters
at once and to analyze the output image data interactively during postprocessing.

The proposed method can be regarded, in some sense, as a generalization of 
the so-called bullet-time method, which is
used in the film industry, where
cameras are placed along a specified path
and pictures are taken sequentially with short time intervals.
This method is used to make a slow motion movie of the target object, which appears
to be taken by a camera moving at an impossibly high speed.
Figure~\ref{fig:concept_01_04_combined}c shows 
a conceptual description
of the bullet-time method where the visualization
is obtained using four different cameras.
Similar to Fig.~\ref{fig:concept_01_04_combined}a,
the output data is a sequence of images, which
is then sent to the analyzer.

In our method,
we use as many in-situ visualizations as possible
and analyze the output image data later in an interactive manner,
as shown in Fig.~\ref{fig:concept_01_04_combined}d.
Any information on the simulated phenomena
is expected to be in the image dataset so
the analyzer can explore the
image space dynamically.

When an output movie from an in-situ visualization 
is compressed to a reasonable size, such as 100~MB,
the total output data size is \textit{only} 100~TB,
even if the number of applied in-situ visualization is one million.
This is still smaller than the size of the raw numerical data 
in exascale simulations, which would be in
the order of PB.

\begin{figure}[htbp]
  \begin{center}
   \includegraphics[width=0.3\columnwidth]{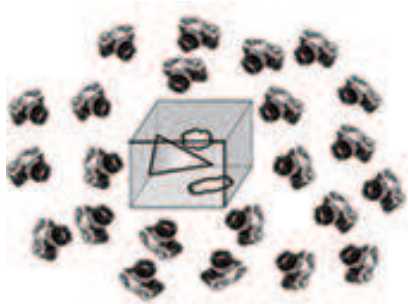}
  \end{center}
  \caption{A cluster of visualization cameras for in-situ visualizations.}
  \label{fig:many_camera}
 \end{figure}

When we select only the camera position among the 
camera parameters,
our proposed method requires that we place 
many cameras inside and outside the simulation region
and that we use them all in the in-situ visualization in parallel, as shown in
Fig.~\ref{fig:many_camera}.
This configuration of cameras reminds us of
the ``3D Dome'' or ``3D Room'' constructed by Kanade et al.~for capturing
human motion in a room-sized space
to construct a virtual reality~\cite{Narayanan1998, Kanade1998}.
Our method can be regarded 
as an extension of the 3D room concept
to scientific visualization.
%An important difference is
%that, cameras 
%can be placed inside the target without affecting it in our in-situ visualization method.
An important difference is that we explore not only in the camera space, 
but also in the visualization space, in our in-situ visualization method.

%-----------------------------------------------------------------------
%-----------------------------------------------------------------------

The output data of our proposed method is a dataset containing many
(thousands to millions) movie files.
When the camera number is sufficiently high,
we can find practically any image from 
any position in the dataset---although 
a large part of the dataset would be useless. 
Therefore, the user can ``rotate'' a visualized object with the mouse,
by extracting an appropriate sequence of image files from the 
movie dataset.

To demonstrate the feasibility of the proposed method,
we performed a batch simulation on a PC cluster system
where $130$ cameras were placed around the simulation region.
Three types of in-situ visualization methods 
were used by all of the cameras during runtime.
The output data comprised $130 \times 3 = 390$ sets of MPEG movie files.
We also developed an interactive movie player
so a sequence of images can be extracted from the dataset.

%==============================
\section{Related Work\label{sec:related work}}
%==============================
In-situ visualization has a long history.
Johnson et al.~\cite{Johnson1999} conducted  a general review of steering simulation and visualization.
Examples of in-situ visualization in peta-scale simulations
and their technical challenges (especially those caused by 
massively parallel processing) are summarized in~\cite{Ma2007,Ma2009}.
They show that in-situ visualization is 
a promising solution for peta and exascale simulations.
A natural extension of the online dynamical control of simulation is the end-to-end approach~\cite{Tu2006,Yu2006},
where even mesh generation, which is usually conducted in the preprocessing stage, is performed on supercomputers.
They developed a steering simulation for seismic wave propagation where the
visualization images were shown in real time.

Huang et al.~developed a steering simulation and visualization framework
for environmental science where dynamical control over the Internet was implemented~\cite{Huang2004}.
Their user interface was constructed on web browsers.
Ellsworth et al.~developed an in-situ visualization system on the Colombia supercomputer for running a weather forecasting model~\cite{Ellsworth2006}.
(Our experimental system described in section~\ref{sec:appseic}
is similar to their system because 
multiple MPEG files are generated by multiple in-situ visualizations.
However, their movies are shown separately in each panel of 
a tiled display system, whereas ours are used as a database.)
Esnard et al.~developed an in-situ visualization system
where the parallel visualization processing can run on a different computer 
system from the simulation computer system~\cite{Esnard2006}.
In their development, special emphasis was placed on 
producing a steering environment with existing simulation codes.

Many in-situ visualizations, with or without steering,
have been designed and developed for particular simulation problems.
However,
general visualization frameworks with high parallel scalability 
have also become available recently.
Whitlock et al.~\cite{Whitlock2011} developed a library \textit{Libsim} that facilitates in-situ visualizations using VisIt, which is
one of the most sophisticated parallel visualization tools available today.
Their paper also contained a concise review of the history and 
the latest status of the in-situ visualization research.
Fabian et al.~\cite{Fabian2011} reported the development of a coprocessing library for \textit{ParaView},
which is another sophisticated parallel visualization tool.
Using that library, it is possible to utilize the various visualization
functions provided by \textit{ParaView} during runtime, 
which is decoupled from the simulation.
The latest case studies of in-situ visualization using \textit{VisIt} and \textit{ParaView} can be found in~\cite{Rivi2012}.

%------------------------------------
\section{Configuration of Viewpoints}
%------------------------------------

One of the key points of our approach to interactive in-situ visualization
is the configuration of multiple cameras.
In an extreme case,
cameras can be placed in a 3-D distribution as densely as possible
inside and around the simulation region.
In our feasibility test in the present study,
we placed only $130$ cameras with a
2-D distribution on a spherical surface 
at a fixed radius from the center of the simulation region,
to ensure that the test was at a moderate level.
All of the cameras pointed at the center.

%---------------------------------------------------------------------------
 \begin{figure}[htbp]
  \begin{center}
   \includegraphics[width=0.3\columnwidth]{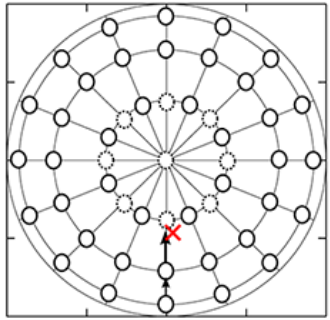}
  \end{center}
  \caption{A spherical camera distribution on a latitude-longitude grid 
  viewed from due north.}
  \label{fig:sphere_model_top}
 \end{figure}
%---------------------------------------------------------------------------

There are only five regular polyhedra so
it is not a straightforward task to place more than 
20 cameras on a spherical surface in a uniform manner.

\begin{figure}[htbp]
  \begin{center}
   \includegraphics[width=0.8\textwidth]{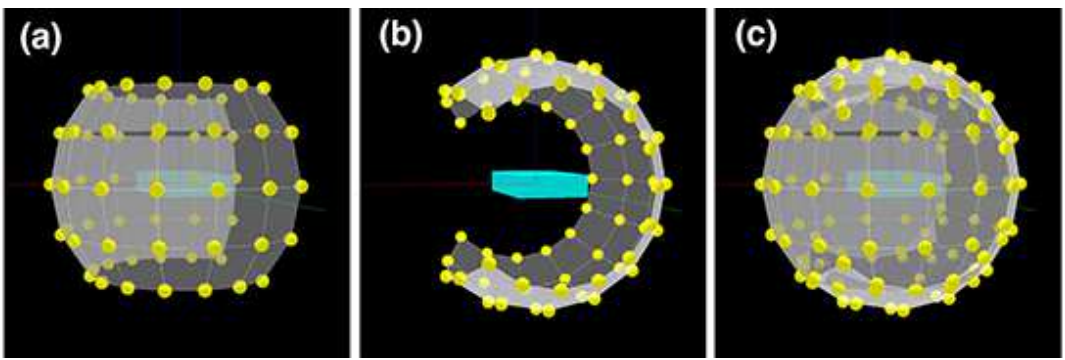}
  \end{center}
  \caption{%
  Cameras (yellow balls) placed on a spherical surface around the simulation 
  region (light blue box) in the experiment described in Section~\ref{sec:appseic}.
  A Yin-Yang grid configuration is used to avoid concentrations of cameras near the poles.
  (a) A set of 65 cameras is placed on the Yin-grid, which is a low-latitude
  subset of the spherical polar coordinates.
  (b) Another set of 65 cameras is placed on the Yang grid, which has the same
  shape as the Yang-grid but with a rotated configuration.
  (c) The Yin- and Yang-grids are combined to cover the entire spherical surface.
  }
  \label{fig:camera_yin-yang_combined}
 \end{figure}

Placing cameras on each grid point of a 
spherical coordinate system, or the so-called latitude-longitude mesh,
does not work well because
it leads to concentrations of cameras around the poles.
Reducing the number of cameras at higher latitudes is not a good idea because
camera motion along a constant longitude
will lead to a lack of viewpoints~(Fig.~\ref{fig:sphere_model_top}).
To avoid this problem,
we use a Yin-Yang grid~\cite{Kageyama2004} where
two congruent component grids, which are a part
of low latitude region of the latitude-longitude mesh,
are combined with a partial overlap
to cover a spherical surface (Fig.~\ref{fig:camera_yin-yang_combined}).

%---------------------------------------------------------------------------
%---------------------------------------------------------------------------

%------------------------------------
\section{Interactive View of Many Movies} \label{sec:ivmm}
%------------------------------------

The camera position is not the only parameter
that can be explored using our method, e.g.,
see the parameter axis in Fig.~\ref{fig:concept_01_04_combined}d.
Different visualization methods and their internal parameters such as the 
isosurface level, slice position, and color functions can 
also be explored using our method.
An ultimate aim of our method is to obtain many (hopefully thousands) different 
visualization parameters from each camera.
In this test, however,
we only used three different visualization parameters.
The full details are provided in the next section.
The total number of output movie files is $390$ $(=130 \hbox{ cameras} \times 3 )$.

These movie files contain a lot of information that 
would be useless without a proper player.
A movie player for PCs usually performs a simple task, i.e.,
it reads a source movie file and shows
the sequence of images contained in the source file on the screen.
Thus, we need a generalized player that can
read multiple ($390$ in the present test and millions in the future) 
source files and display a sequence of images derived from them.
The play should be interactive, i.e.,
the source should be instantly and smoothly changed based on 
a user's input such as a mouse motion or keystroke.

%------------------------------------------------------------------------------------------------
\begin{figure}[htbp]
\begin{center}
\includegraphics[width=0.6\columnwidth]{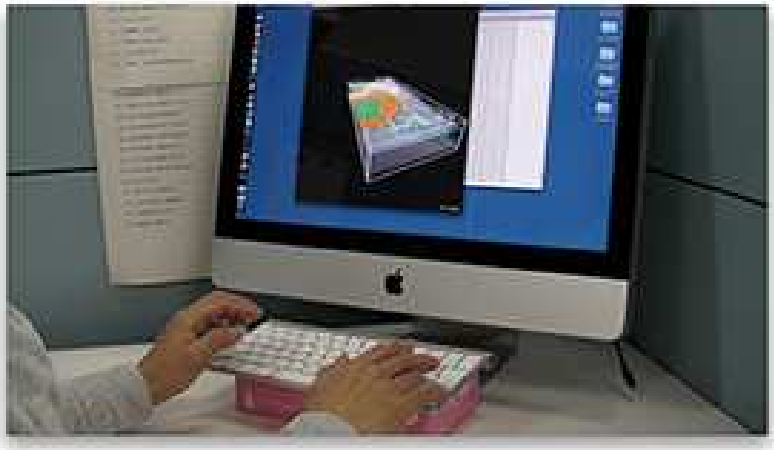}
\end{center}
 \caption{The special movie player developed in this study.
 Multiple movie files are loaded into the player and
 sequences of images from different movie files can be retrieved and shown on the screen.}
 \label{fig:demo}
\end{figure}
%------------------------------------------------------------------------------------------------

We have developed an interactive movie player using the OpenCV library.
A snapshot is shown in Fig.~\ref{fig:demo}.
Our movie player can retrieve a sequence of images from different movie 
files and display them on the window as a single movie.
The player we developed has three functions:
(i) to stop and play the motion; 
(ii) to change the source movie;
and (iii) to zoom in/out of a scene.

The zoom in/out function was implemented using the image magnification function provided
by OpenCV.
If we placed many more cameras 
in a fully 3-D distribution,
we would be able to
perform the zoom in/out function by changing the source cameras.

The aim of the method proposed in this paper is to cover
the whole visualization parameter space by applying the in-situ
visualizations at once.
As shown in Fig.~\ref{fig:concept_01_04_combined}d,
the visualization parameter space is covered not only by
the camera position but also by the applied visualization methods and their
internal parameters.
In this experiment, we have used just one visualization
method (volume rendering) with three different types of target data.
In exascale applications, we will apply 
$10^3$ different visualization methods and parameters for each camera.
The movie player that we have developed in this study
can extract any image sequence from different movie files in the loaded movie dataset, so
there are no technical restrictions on interactive changes in the 
visualization parameters, even in the present version of the movie player.
For example,  we can smoothly change the isosurface level by a keystroke or 
mouse drag if we have isosurface visualizations for each camera
with sufficient different isosurface levels.

As for the volume rendering, it is difficult 
to specify a proper range of the transfer function beforehand.
We would not, however, take a special care for the range setting,
by accepting the possibility that large part of the
volume rendering images is useless 
due to unsuitable transfer functions,
as we have already accepted the possibility that
many cameras are taking images from useless angles
in the proposed method.
A point in this method is to find
valuable information afterwards in a large amount of movie database.

%------------------------------------
\section{Application to Seismic Wave Propagation Simulation\label{sec:appseic}}
%------------------------------------
In our feasibility study,
we performed a seismic wave propagation simulation
using our proposed in-situ visualization method.
The simulation and visualization codes were 
developed by Furumura and Chen.
The details of the codes are described in
their paper~\cite{Furumura2004},
so we briefly summarize only the key features 
related to our in-situ visualization experiment.

This simulation solves the time development of 
seismic waves in Cartesian coordinates $(x,y,z)$.
The basic variables are the displacement 
field ${\bf u}(x,y,z)$ of a viscoelastic medium
and its velocity ${\bf v}(x,y,z)$.
The viscoelastic equations of motion are solved for them
using a high-order finite difference method.
A pseudo-spectral method can also be selected
in their code if the spatial accuracy is critically important.
However, we used only the finite difference method in this test.

One-dimensional domain decomposition is applied
on the $z$ (vertical) axis 
for the parallel simulation.
The rectangular simulation region is divided into horizontal slates.
This code has good, almost linear, scaling up to more than 1000
processors in the 
Earth Simulator supercomputer~\cite{Furumura2004}.
Hybrid parallelization is used with MPI and OpenMP.

Furumura and Chen also developed visualization tools for
their seismic wave propagation simulation.
These tools were 
derived from a highly optimized parallel visualization package~\cite{Chen2003}, which was 
developed in the GeoFEM framework~\cite{Okuda2003}
for the Earth Simulator supercomputer.
The original visualization package was designed to use unstructured cell data.
In this visualization tool for seismic wave propagation, however,
they were applied to a Cartesian structured mesh.
Various visualization methods based on parallel ray castings were implemented
in the original package of GeoFEM.
In this test, we used only one type of visualization method, i.e., 
volume rendering for scalar fields.

Although the visualization tools used in this experiment
originate from the GeoFEM package for unstructured cell data, 
the source code has been  converted into a version 
that can be applied directly to Cartesian structured mesh data.
Therefore, we could apply the visualization to the
simulation data without converting the data format.

More sophisticated visualization packages
such as VisIt and ParaView are now available.
These modern packages provide highly scalable parallel rendering 
that can be integrated with the simulation code, so they would be a much better choice 
for the rendering engine of the in-situ visualization method proposed in this paper for
general simulations.
However, we used the GeoFEM-derived visualization tool in this prototype experiment
simply because it was strongly coupled to the seismic wave propagation simulation.
We will implement our method with ParaView in the future.

Seismic waves have two modes: s-waves (shear mode) and p-waves (compressional mode).
We defined two scalar fields $\phi_s\equiv |\nabla\times{\bf v}|$ (for s-wave)
and $\phi_p\equiv \nabla\cdot{\bf v}$ (for p-wave), and
we applied 
the in-situ volume rendering visualization
to the two fields $\phi_p$ and $\phi_s$.

We performed test runs on a PC cluster system where the
simulation grid size was $256\times 512 \times 160$.
The number of MPI processes was 40.
As mentioned above,
130 cameras were placed spherically on the Yin-Yang grid points, i.e., 65 for Yin and 65 for Yang.
Each camera performed three visualization tasks: volume rendering
for the s-wave, the p-wave, and both of them.
Visualization snapshots were captured 60 times during one simulation task.
The images were saved in the PPM format and the size of each image
was $2.5~$MB.
The total size of all the images was $57\hbox{ GB} \ (=
2.5 \hbox{ MB} \times 130 \hbox{ cameras} 
\times 3 \hbox{ tasks} \times 60 \hbox{ frames}$).
After the simulation task,
each set of sequential images captured by a single camera
was combined into a movie file in the MPEG format
using the FFmpeg library.
(We will implement runtime compression in the future.)
The compression rate was about $3~\%$;
the total PPM image set of $57$~GB was 
converted into MPEG files with a total size of $1.7$~GB.

If we would save the raw numerical data from $\phi_s$ and $\phi_p$ with
single-point precision for postprocessing,
it would amount 
to $10\hbox{ GB}\  (=256\times 512 \times 160 \hbox{ grids} \times 4 \hbox{ B} \times 2 \times 60 \hbox{ snapshots}$).
Our movie data ($1.7$~GB) was an order of magnitude smaller than this.
This gap will increase further in larger scale simulations.

The total number of movie files produced during this experiment
was $390$\ $(=130 \hbox{ cameras} \times 3 \hbox{ tasks})$.
The 390 MPEG files of $1.7$~GB 
were transferred to a PC, as shown in Fig.~\ref{fig:demo}, 
and they were loaded into the
interactive movie player described in Section~\ref{sec:ivmm}.

%--------------------------------------------------------------------------------------------------
\begin{figure}[htbp]
\begin{center}
\includegraphics[height=0.8\textheight]{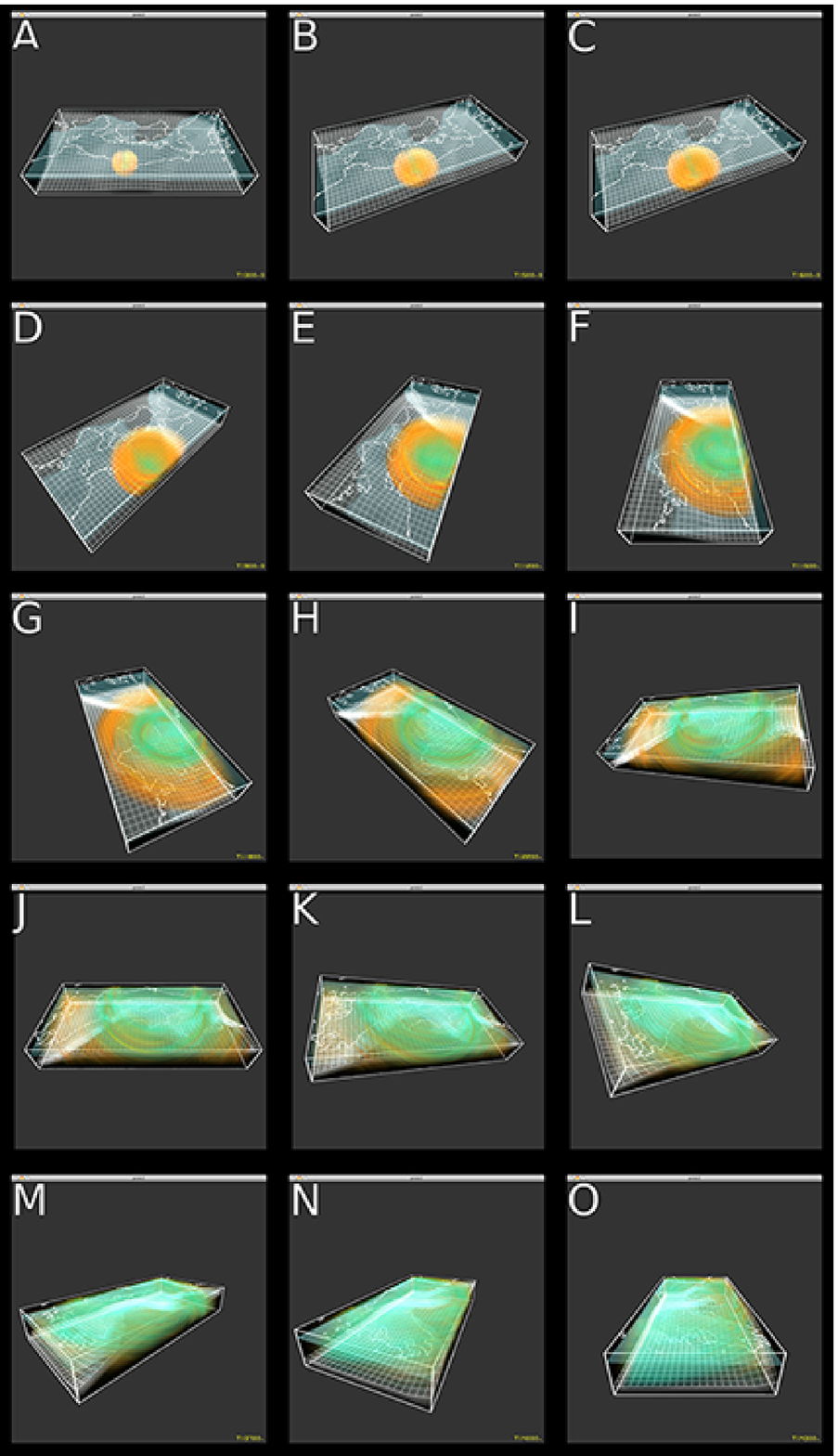}
\end{center}
 \caption{A snapshot sequence of the movie player.
 The user can rotate the viewing angle by typing the arrow 
 keys (right/left/up/down) on the keyboard,
 while the p-wave (orange) and s-wave (green) are propagated.
 Note that this is not a steering simulation or a 
 postprocessing visualization in the usual sense.
 The simulation has already been performed as a batch job.
 }
 \label{fig:omatomego18-2}
\end{figure}
%--------------------------------------------------------------------------------------------------

Figure~\ref{fig:omatomego18-2} shows a sequence of
snapshots with an interactive view of the in-situ visualization movie data.
The propagation of the p-wave (orange) and s-wave (green) can be 
observed in the movie displayed in a PC window.
Typing a key on the keyboard sends a signal to the movie player, which 
changes the source movie file (or changes the camera angle).
The user can observe these phenomena from any one of the spherically
distributed 130 cameras
shown in Fig.~\ref{fig:camera_yin-yang_combined}.
Image retrieval from the movie dataset is smooth and fast.
The user can change the camera position
while the seismic wave propagates in the window.
Despite the relatively small number of cameras,
the change of view is so smooth that
the user might think that this is a normal postprocessing visualization.

%------------------------------------------------------------------------------------
\begin{figure}[htbp]
\begin{center}
\includegraphics[width=0.7\columnwidth]{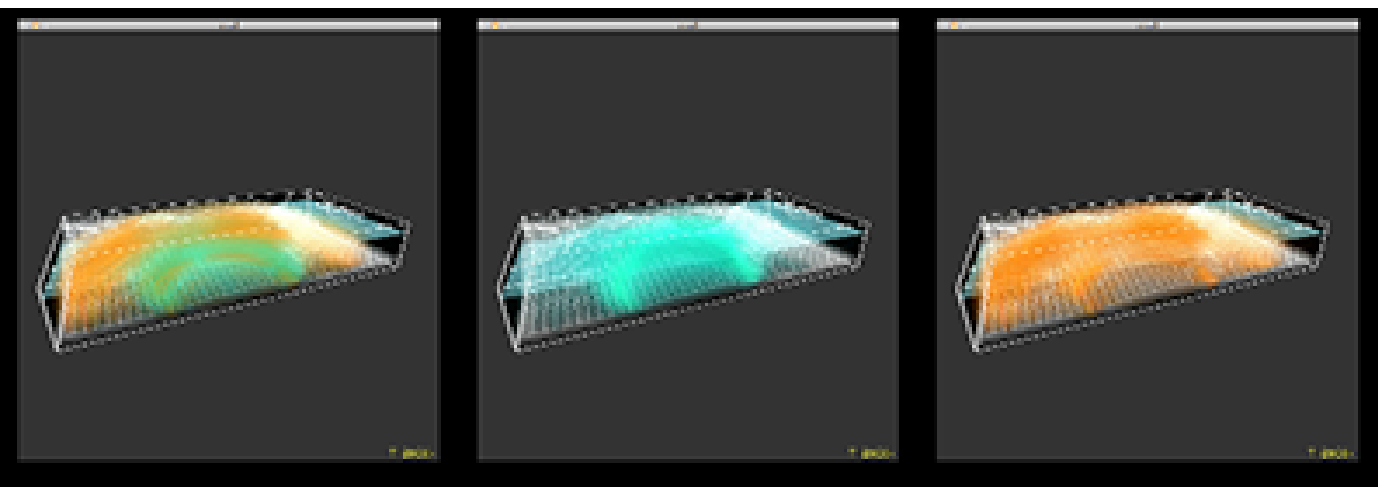}
\end{center}
 \caption{Interactive control of visualization parameters other
 than the camera position.
 The user can change the visualization task, the
 volume renderings of the p-wave (right panel) and the
 s-wave (left panel), or both of them (left panel),
 by typing a key on the keyboard while wave
 propagation is observed.
 }
 \label{fig:p_s_ps}
\end{figure}
%------------------------------------------------------------------------------------

Figure~\ref{fig:p_s_ps} shows snapshots of the interactive switching of visualization tasks.
At the beginning, the p-wave and s-wave are visualized in 
the window.
At this point, the movie player is retrieving an image sequence from a movie file, which
contains images of the p-wave and s-wave.
When the viewer types a key,
the movie player changes the source file and starts to
retrieve image sequence from the new source, which may contain
only the s-wave images.
Thus, the window will show wave propagation.
The user does not experience any delay during this switch.
Another keystroke produces an instant switch in
the visualization task to focus on p-wave propagation.

\begin{figure}[htbp]
\begin{center}
\includegraphics[width=0.7\columnwidth]{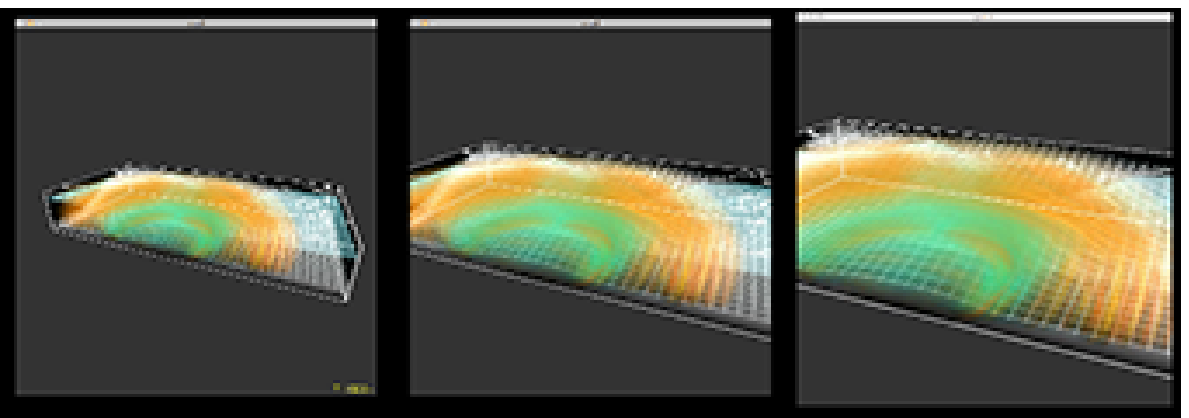}
\end{center}
 \caption{Close-up view of the data.
 Zooming was implemented by image
 processing. In this case, the magnifications were
 of 1.0 (left panel), 1.5 (middle), and 2.0 (right).
 If other cameras were placed closer to the target, the user
 could change the source to another camera, similar to
 the camera rotation shown in Fig.~\ref{fig:omatomego18-2}.
 }
 \label{fig:zoom_scene}
\end{figure}

Figure~\ref{fig:zoom_scene} shows the zoom function available in the movie player.
As mentioned in Section~\ref{sec:ivmm},
we will place a high density of cameras in the most uniform 3-D configuration possible
in our final implementation in the future.
Thus, if we need to focus on a small area in the simulation region,
our interactive movie player will extract other image sequences captured by
a camera located closer to the target region.
In the current test, however,
we implemented the zoom function by magnifying images captured by a
fixed camera.

%-------------------------------------------------------------------------------------------
%-------------------------------------------------------------------------------------------

%-----------------------------------------------
\section{Concluding Remarks}
%-----------------------------------------------
In this paper, we have proposed a new in-situ visualization method for 
exascale simulations.
This method can generate numerous (hopefully millions) in-situ visualizations simultaneously
based on thousands of different visualizations captured by many
(possibly thousands) different cameras.
The output of this simulation strategy is movies, rather than numbers.

When a simulation produces one million movies,
the total output data size is only 100~TB
if each movie is compressed to 100~MB.
It should be noted that a size of 100~TB
will be quite small in the coming exascale era.
The output data would be much larger if a user tries to save raw numerical data with the same temporal resolution
for standard postprocessing visualization.

In our experiment, each in-situ visualization movie
was compressed to $2.5~$MB
for each visualization parameter (including the camera position) and visualization method.
The pixel size of a movie was $920 \times 896 $ and the frame number was 60.
The frame number of 60 would need to be increased in future exascale applications
and each image would require finer resolution
because the phenomena simulated in future simulations are expected to have 
higher complexity.
However, we would not have to produce visualizations of ``ultra-fine'' resolution movies, even if ``ultra-fine'' simulations are performed.
This is because if we need to analyze a fine structure in a simulation region,
there would be a camera close to that spot and visualization movies could be captured
at a moderate resolution,
e.g., $1600 \times 1200$ pixels (UXGA), 
which would provide the necessary information.
Therefore, a movie file size of $100$ MB 
would be a reasonable estimate.

An output of one million movies can be analyzed during postprocessing
using our proposed method.
Our specially designed movie player 
reads the one million movie files and displays
a sequence of images in a window.
After obtaining an appropriate image sequence from 
different movie files,
we can ``rotate'' the visualization objects effectively
while the dynamic phenomena are displayed in the window.
We can also change the visualization method
by using different internal parameters, provided the corresponding in-situ visualizations have 
been applied during the simulation task.

To demonstrate the feasibility of our proposed method,
we performed a simulation
where 390 in-situ visualizations were generated.
We also developed a movie player that reads 390 movie files
and displays a movie in a PC window.
Using this player, we confirmed that
it was possible to ``rotate'' a visualized object interactively
and to change the visualization task in real time
by a single keystroke while the movie was playing.

Clearly, the size of our experiment was
far from the exascale range.
The major difference is in the size of the output dataset.
In our experiment, the number of the movie files was only 390,
whereas that would be $10^6$ in exascale tasks because
1000 cameras will be placed in and around the simulation region,
and each camera will capture 1000 different visualizations.
Since a set of 
visualizations by different cameras is an embarrassingly parallel problem,
the scalability for the number of cameras does not matter
as long as plenty number of computer nodes are available.

The size of each movie file will not increase dramatically 
even with the development of 
exascale applications, provided that researchers view the movies on 
monitor windows with 2k to 4k pixel widths.
In our experiment, reading 390 movie files (1.7 GB) required
1.88 s when using an Apple iMac (SATA 7200 rpm) and 
1.42 s with an HP Z800 (SATA 7200 rpm).
Extrapolating the read time in a linear manner,
it will take $3600\sim 4800$ s  
if we load $10^6$ movie files in the memory and
the memory size required is $4.4$TB.
We cannot predict the main memory size available and the I/O 
bandwidth of storage devices in the exascale era, but
it is possible that we will need to take special care when handling such a large dataset.
A possible solution is a prefetch mechanism, which would allow the dynamic loading
of part of the movie dataset
that is expected to be required by the user's next request.
We will implement this prefetch mechanism in our movie player.

Our proposed method can be summarized 
as follows:
(i) identify the potential ranges for the visualization and the camera parameters
that will be of interest to the user;
(ii) discretize the parameter subspace at as fine a level as possible;
(iii) apply in-situ visualizations to all of the discrete 
parameters (the white balls in Fig.~\ref{fig:concept_01_04_combined}d)
and save the output movies; and
(iv) explore the movie data space using an interactive movie player. 
Any information required for the analysis will be present in the movie dataset.

In conclusion, our experiment 
suggests that this method of in-situ visualization
with interactive viewing will be practical for peta and exascale supercomputing.

%------------------------------------
 \section*{Acknowledgements}
%------------------------------------
We thank Professor Furumura for providing 
the simulation and volume rendering codes for seismic wave propagation.
This work was supported by
Grant-in-Aid for Scientific Research (KAKENHI) 23340128 
and the Takahashi Industrial and Economic Research Foundation.

%
%\bibliographystyle{elsarticle-num}
%\bibliography{121128}
%

\end{document}